\documentclass[10pt,conference]{IEEEtran}
\IEEEoverridecommandlockouts
\usepackage{cite}
\usepackage{amsmath,amssymb,amsfonts}
\usepackage{algorithmic}
\usepackage{graphicx}
\usepackage{textcomp}
\usepackage{xcolor}
\usepackage{subfig}
\usepackage{bm}
\usepackage{algorithm}
\usepackage{subcaption}
\usepackage{multirow}

\usepackage{tabularx}
\usepackage{booktabs}
\usepackage{mathrsfs}
\def\BibTeX{{\rm B\kern-.05em{\sc i\kern-.025em b}\kern-.08em
    T\kern-.1667em\lower.7ex\hbox{E}\kern-.125emX}}
\begin{document}

\title{Semantic Communication for Efficient Point Cloud Transmission}

\author{\IEEEauthorblockN{Shangzhuo Xie\IEEEauthorrefmark{1}, Qianqian Yang\IEEEauthorrefmark{1}, Yuyi Sun\IEEEauthorrefmark{2}, Tianxiao Han\IEEEauthorrefmark{1}, Zhaohui Yang\IEEEauthorrefmark{1}, Zhiguo Shi\IEEEauthorrefmark{1}}
\IEEEauthorblockA{ \IEEEauthorrefmark{1}College of Information Science and Electronic Engineering, Zhejiang University, Hangzhou 310007, China \\
\IEEEauthorrefmark{2}College of Information Science and Technology, Hangzhou Normal University, Hangzhou 311121, China \\
Email: 
\{3200104418,qianqianyang20,Tianxiao Han,yang\_zhaohui,shizg\}@zju.edu.cn,   yuyisun@hznu.edu.cn
}
}

\maketitle

\begin{abstract}
As three-dimensional acquisition technologies like LiDAR cameras advance, the need for efficient transmission of 3D point clouds is becoming increasingly important. In this paper, we present a novel semantic communication (SemCom) approach for efficient 3D point cloud transmission. Different from existing methods that rely on downsampling and feature extraction for compression, our approach utilizes a parallel structure to separately extract both global and local information from point clouds. This system is composed of five key components: local semantic encoder, global semantic encoder, channel encoder, channel decoder, and semantic decoder. Our numerical results indicate that this approach surpasses both the traditional Octree compression methodology and alternative deep learning-based strategies in terms of reconstruction quality. Moreover, our system is capable of achieving high-quality point cloud reconstruction under adverse channel conditions, specifically maintaining a reconstruction quality of over 37dB even with severe channel noise. 
\end{abstract}


\section{Introduction}

The 3D point cloud is widely used for representing three-dimensional data, with applications in autonomous driving and virtual reality \cite{application, air}. Transmitting 3D point clouds wirelessly imposes strict requirements on bandwidth, latency, and reconstruction quality. Over the years, various point cloud encoding methods have been developed, including geometry-based \cite{GPCCTEST}, voxelization-based \cite{mesh}, 3D convolution-based \cite{3Dconv}, and nerf-based \cite{neural} approaches. However, relying solely on source coding can result in `cliff effect' where decoding performance drastically deteriorates under harsh channel conditions. Recently, emerging semantic communication (SemCom) approaches address this issue by integrating source and channel coding, optimizing data transmission for tasks rather than precise bit recovery, which has demonstrated improved transmission efficiency \cite{loT}.

SemCom has been widely applied for the transmission of images, videos, and speech \cite{twostage}, but its use in point clouds remains limited. Notable efforts include the Point Cloud-based SemCom System (PCSC) \cite{PCSC}, which employs the Voxception-ResNet (VRN) network for joint source-channel coding of point clouds, achieving improved transmission efficiency compared to traditional methods. The authors in \cite{SEPT} utilized Point Transformer to build a SemCom system, eliminating cliff effects and offering improved compression performance under AWGN channels. However, point clouds are characterized by large volumes and complex geometric features. These methods do not effectively extract rich and multi-level semantic information within point cloud data. 

Existing work on semantic communication (SemCom) for image transmission has validated that multi-level semantic information can help preserve semantics in images, leading to better transmission performance, especially under harsh channel conditions \cite{multi}. Promising results have also been seen in point cloud processing, where this approach successfully extracts global and local semantic information and supports various downstream tasks without complex networks \cite{multipoint}.

Motivated by these works, this paper presents a point cloud semantic communication (SemCom) system framework based on multi-level semantic information extraction. The framework utilizes a point-based neural network to extract local semantic information from point cloud segmentation patches and a graph-based neural network to extract global semantic information from point cloud projection images. To further enhance transmission quality, a hierarchical transmission strategy is implemented: lossless transmission, such as PBRL-LDPC with HARQ, is used for transmitting global semantic information and patch centroid coordinates, which have a smaller data volume. In contrast, lossy transmission is employed for local semantic information through a joint source-channel coding codec. Our simulation on the ModelNet40 dataset demonstrated that our network can maintain a reconstruction quality of around 37 dB even under extremely poor channel conditions, representing an improvement of more than 15\% compared to existing method.



\begin{figure*}[tbp]
\includegraphics[width=1\textwidth]{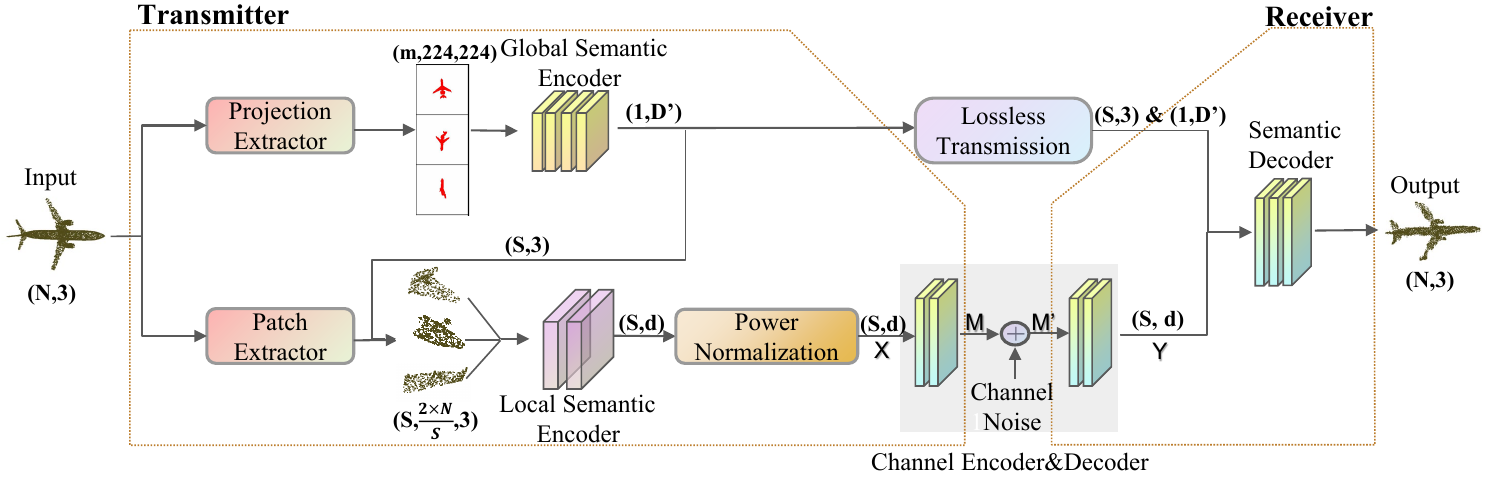} 
\centering 
\caption{The overall framework of our point cloud semantic transmission system}
\label{model_total}
\end{figure*}

\section{System Model}
In this section, we introduce the system model for a SemCom system aimed at efficient point cloud transmission and outline the performance metrics used to evaluate transmission quality.
\subsection{SemCom for Point Clouds}
The main framework of our point cloud transmission system is illustrated in Fig.~\ref{model_total}. The point cloud $\bm{P}$ contains only three-dimensional coordinates, represented as $\boldsymbol{P} = (N, 3)$. At the transmitter, $\bm{P}$ is first processed by the patch extractor, producing $S$ patches $\boldsymbol{Pc} = \{Pc_1, Pc_2, ..., Pc_s\}$, each containing $2 \times N/S$ points. Corresponding centroid coordinates $\boldsymbol{Px} = \{Px_1, Px_2, ..., Px_s\}$ are also obtained. The patches $\boldsymbol{Pc}$ are then sent to the local semantic encoder, generating local semantic information $\boldsymbol{L}$, a combination of $d$-dimensional local semantic vactor from each patch. Simultaneously, $\boldsymbol{P}$ is processed by a projection extractor, producing $m$ projection images $\boldsymbol{Pr} = \{Pr_1, Pr_2, ..., Pr_m\}$, each with a size of $224 \times 224$. These images are processed by the global semantic encoder to generate global semantic information $\boldsymbol{G}$, which contains $D'$-dimensional global features of $\boldsymbol{P}$.

After that, global information $\boldsymbol{G}$ and centroid coordinates $\boldsymbol{Px}$ are transmitted losslessly. For local information $\boldsymbol{L}$, power normalization is applied, resulting in $\boldsymbol{X} = (S, d)$, which is then encoded and mapped by the channel encoder, producing $\boldsymbol{M} = (S, 2 \times d)$. $\boldsymbol{M}$ is then mapped into a complex vector $\bm{\hat{M}} \in \bm{C}^{S \times d}$. The received signal for the decoder is given by:
\begin{equation}
\bm{\hat{M'}} = \bm{\hat{M}} + \mathbf{n}
\end{equation}
$\mathbf{n}$ represents the Gaussian noise $\mathcal{C}\mathcal{N}(0, \sigma^2\bm{I})$ with the average power of 1.

The receiver then reshape $\bm{\hat{M'}}$ into $\bm{M'} \in \bm{R}^{S \times 2 \times d}$, and input to the channel decoder to obtain new local information $\bm{Y}=(S,d)$. The global information, the decoded local information and centroid coordinates are then fed to the semantic decoder to reconstruct point cloud $\bm{P'} = (N,3)$.

\subsection{Performance Metrics}
To evaluate transmission performance, we use two commonly applied peak signal-to-noise ratio (PSNR) metrics, D1 and D2 \cite{PSNR}. Assume $\bm{A}$ and $\bm{B}$ are the original and transmitted point clouds, respectively. 


D1 calculates the point-to-point distance between $\bm{A}$ and $\bm{B}$. The compression error $e^{c2c}_{\bm{A,B}}$ is given by: \begin{equation}
e^{c2c}_{\bm{A,B}} = \frac{1}{N_A} \sum_{a_j\in\bm{A}}||E(i,j)||^{2}_{2}
\label{ec2c}
\end{equation}
where $E(i, j)$ represents the geometric distance between each point $a_j$ in $\bm{A}$ and its nearest point $b_i$ in $\bm{B}$.

For the point-to-plane distance between $\bm{A}$ and $\bm{B}$, we construct a plane using the normal vector method described in \cite{PSNR} and project $E(i, j)$ onto this plane to obtain the point-to-plane compression error $e^{c2p}_{\bm{A,B}}$: 
\begin{equation}
e^{c2p}_{\bm{A,B}} = \frac{1}{N_A} \sum_{a_j\in\bm{A}}||E’(i,j)||^{2}_{2}
\label{ec2c}
\end{equation}
where $E'(i, j)$ is the projection of $E(i, j)$ onto the plane.

Finally, the PSNR is calculated using $e_{\bm{A,B}}$: 
\begin{equation}
PSNR_{\bm{A,B}} = 10\log_{10}\left(\frac{p^2}{e_{\bm{A,B}}}\right)
\label{PSNR}
\end{equation} 
where $p$ is defined as the diagonal distance of the point cloud bounding box. D1 aligns with intuitive metrics for point cloud changes and transmission quality, while D2, better suited for point clouds with structural surface features, offers a more perceptual quality evaluation.

\section{Proposed Method}

In section \uppercase\expandafter{\romannumeral3}, we discuss the detail of the encoder and decoder design in our transmission system.

\subsection{Patch Extractor and Local Semantic Encoder}

The detailed structure is illustrated in Fig.~\ref{model_part}a. The process begins with capturing a patch of the point cloud, which is then fed into the local semantic encoder. For patch acquisition, we use the farthest point sampling (FPS) \cite{FPS} algorithm to sample the input point cloud $\bm{P}(N,3)$, ensuring the patch centroids $\bm{C}(S,3)$ are as evenly distributed as possible. Next, the k-nearest neighbors (kNN) \cite{KNN} algorithm is applied to each centroid, sampling $K$ points around each centroid to form a patch. As a result, $S$ patches are generated, each containing $K$ points. To maximize point coverage, oversampling is incorporated, potentially sampling every point in the point cloud. The relationship between $S$ and $K$ is defined as:

\begin{equation}
S \times K = 2 \times N
\label{SK}
\end{equation}

To mitigate the impact of patch offset on the local semantic encoder, we subtract the centroid coordinates from all points within each patch. Each patch $(K, 3)$ is then individually processed by the local semantic encoder. The encoder comprises a PointNet++ \cite{PointNet++} layer followed by a PointNet layer. The PointNet++ layer includes a downsampling layer (FPS), a grouping layer (KNN), and a PointNet layer. The PointNet layer consists of three MultiLayer Perceptron (MLP) layers and a max-pooling layer. In the downsampling layer, the number of sampling points is set to $K/2$, while the grouping layer contains $K/4$ collection points. The final PointNet layer extracts features with a dimensionality of $d$. As a result, the dimensionality of the information for each patch transforms as follows: 

\begin{flalign*}
&\text{FPS}[({K,3})] \to \text{KNN}[({K/2,3})] \to \text{PointNet}[(K/2,K/4,3)] &\\
&\to \text{PointNet}[(K/2,128)] \to (1,d) &
\end{flalign*}

We obtain the local semantic information $\bm{L}(S, d)$ of $S$ patches. It is worth noting that since our local semantic encoder extracts information for each patch, its network structure is lightweight.

\subsection{Projection Extractor and Global Semantic Encoder}

\begin{figure*}[!t]
  \centering
  \subfloat[]{\includegraphics[scale=0.5]{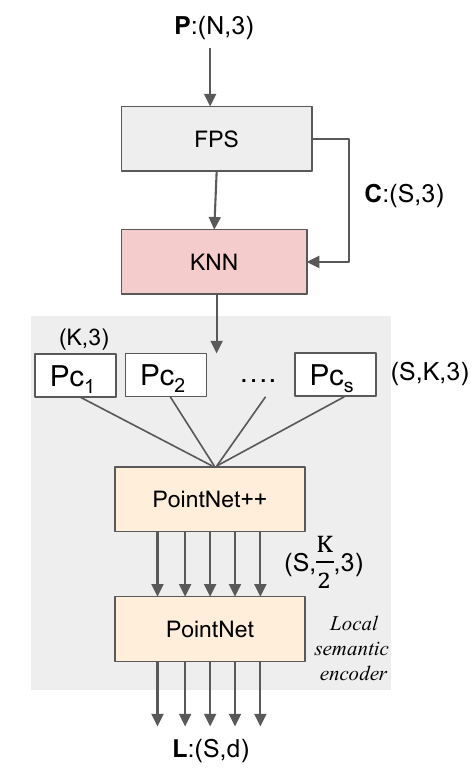}}
  \subfloat[]{\includegraphics[scale=0.5]{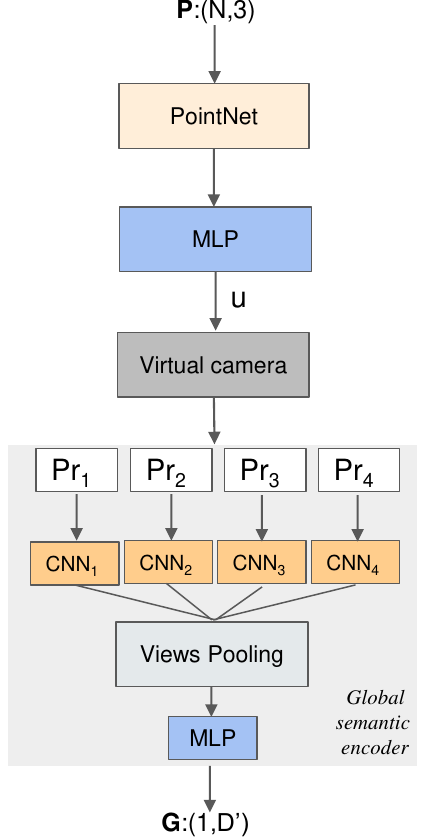}}
  \subfloat[]{\includegraphics[scale=0.5]{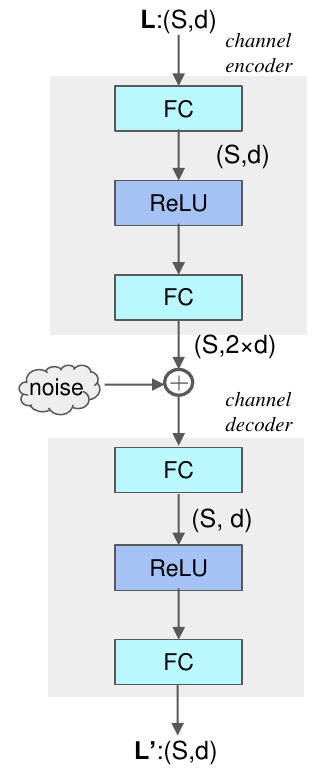}} 
  \subfloat[]{\includegraphics[scale=0.5]{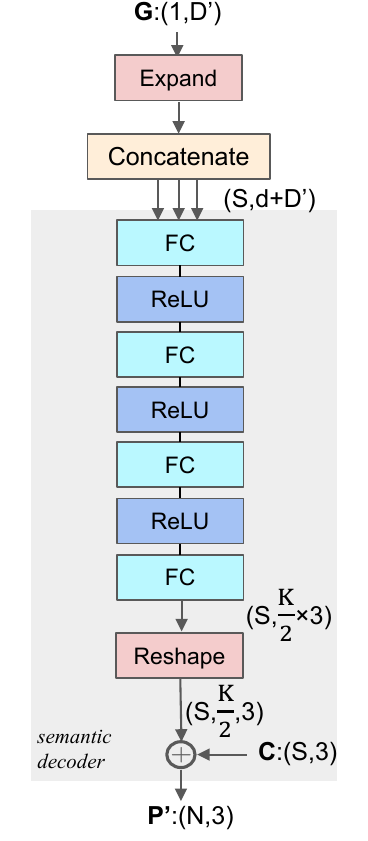}} 
  \caption{The specific structure of each module, from left to right, is the local semantic encoder, global semantic encoder, channel encoder and decoder, semantic decoder. }\label{model_part}
\end{figure*}

The global semantic encoder's structure is illustrated in Fig.~\ref{model_part}b. To extract global semantic information, we first generate the necessary projection maps. Traditionally, this can be done in two ways: 1) Surround the 3D model with a regular icosahedron, placing a virtual camera at each plane's center, and rotate it to capture 80 projection maps. 2) Capture 12 projection maps by taking photos every 30 degrees around the z-axis. To keep our model lightweight while effectively extracting global semantics, we follow the approach from MVTN \cite{MVTN}. The point cloud $\bm{P} (N, 3)$ is processed by PointNet and an MLP to obtain the virtual camera parameters $\bm{u}$, which include azimuth, elevation, and distance information. Using these virtual cameras, we generate four projection images, each with a pixel size of (224, 224), where the point cloud is rendered in red on a white background.

The projection maps are then fed into four Convolutional Neural Networks (CNNs) with shared parameters to capture global information. A views pooling layer consolidates the 512-dimensional global data extracted by the CNNs. Finally, a three-layer MLP reduces dimensionality and extracts global semantic information $\bm{G}(1,D')$.

\subsection{Channel Encoder and Decoder}

The model architecture is shown in Fig.~\ref{model_part}c. We input the power-normalized semantic data $\bm{L} (S, d)$ into a channel encoder consisting of two fully connected (FC) layers, which scales the data to $(S, 2 \times d)$. Before transmission, the data is subjected to noise interference. At the receiver, the noise-distorted data $(S, 2 \times d)$ is processed by a channel decoder, also containing two fully connected layers, to correct errors and regenerate the local semantic information $\bm{L'} (S, d)$ required by the semantic decoder.

\subsection{Lossless Transmission}

According to \cite{lossless transmission}, when the channel code rate is set at $1/2$ and BPSK modulation is adopted, it is feasible to achieve a success rate of $p=0.9$ for transmissions at a Signal-to-Noise Ratio (SNR) of 0, provided the block length exceeds 128. The number of successful transmissions follows a geometric distribution with probability $p$, implying that the expected number of lossless transmissions is $1/p$. Therefore, the required symbol count for lossless transmission under these parameters can be determined as follows:
\begin{equation}
\text{{Channel symbol use}} = \frac{1}{p} \times \left( \frac{2 \times \text{{Bit use}}}{\mathcal{C}} \right)
\label{use cal}
\end{equation}

$\mathcal{C}$ represents the channel capacity. In this paper, we can calculate it as follows:
\begin{equation}
\mathcal{C} = \log_2(1 + 10^{(\frac{{SNR_{dB}}}{10})}) 
\end{equation}

In subsequent experiments, we directly package the center of mass coordinates $\bm{C}(S, 3)$ and global information $\bm{G}(1, D’)$ for lossless transmission, calculating the model's compression rate using Eq.~\ref{use cal}. It should be noted that this equation provides the maximum transmission volume required for lossless transmission. When channel conditions improve, the actual transmission volume will be lower than the calculated value.

\subsection{Semantic Decoder}

As shown in Fig.~\ref{model_part}d, the semantic decoder begins by expanding $\bm{G}(1,D')$ into $\bm{G'}(S,D')$, reflecting that different patches of the same point cloud share identical global semantic information. We then concatenate local and global semantic information, resulting in a total feature dimension of $d+D'$ for each patch. To maintain a lightweight decoder structure, patches are processed separately. The upsampling layer, built using a fully connected layer, outputs data of size $K/2 \times 3$, which is reshaped into patches of size $(K/2,3)$, consisting of $K/2$ points. According to Eq.~\ref{SK}, after upsampling each patch, the total number of reconstructed points $N$ matches the original input point cloud.

Next, we align the patches based on their centroid coordinates and merge them to obtain the reconstructed point cloud $\bm{P'}$. In our model, we calculate the Chamfer distance (CD) \cite{cd} between $\bm{P}$ and $\bm{P'}$ as the loss function, defined as follows:
\begin{equation}
\begin{split}
CD(\bm{P}, \bm{P'}) = \frac{1}{N} \Bigg[ & \sum_{p\in\bm{P}}\min_{p'\in\bm{P'}}||p-p'||_2^2 \\
& + \sum_{p'\in\bm{P'}}\min_{p\in\bm{P}}||p'-p||_2^2 \Bigg]
\end{split}
\end{equation}

\subsection{Two-Stage Training}

In our model, we employ a two-stage training strategy to enhance adaptability to various channel conditions, a method proven effective in SemCom systems \cite{twostage}. In the first stage, the model excludes the channel encoder and decoder, focusing on training the local semantic encoder, global semantic encoder, and semantic decoder in a noise-free environment.

In the second stage, we introduce the channel encoder and decoder, loading the pre-trained local and global semantic encoders, as well as the semantic decoder. The parameters of the local and global semantic encoders are fixed since they have already demonstrated effective information extraction without noise interference during the first stage. We then add noise to train the channel encoder, channel decoder, and semantic decoder, ensuring they work together to mitigate noise.

This strategy allows us to quickly adapt the model to different channel conditions by directly applying second-stage training to the pre-trained model, significantly reducing training time.

\section{Experimental Results}

\begin{figure}[!t]
  \centering
  \subfloat[D1]{\includegraphics[width=0.9\linewidth]{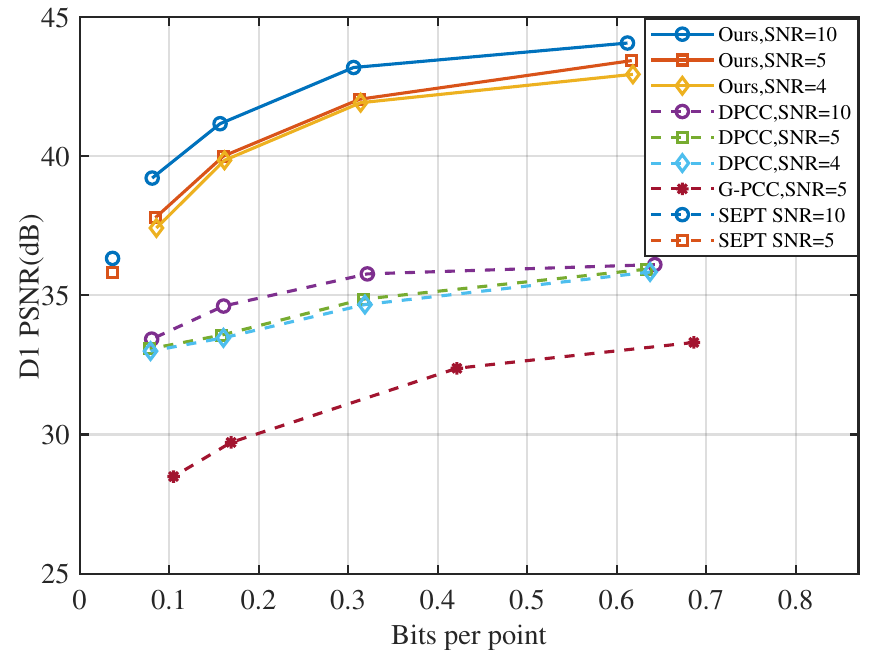}}\\
  \subfloat[D2]{\includegraphics[width=0.9\linewidth]{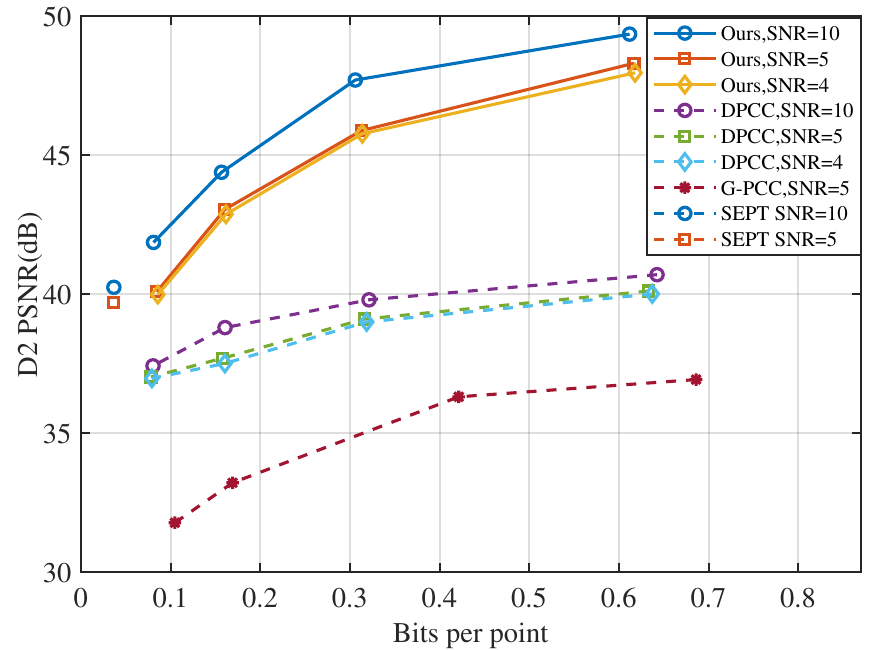}}
  \caption{The reconstruction performance of the model under different compression performance, we set SNR=$\{10,5,4\}$. At the same time, we show the performance of SEPT when bottleneck size=$300$.}
  \label{BPP}
\end{figure}

\begin{figure}[htbp]
\centering
\includegraphics[width=0.9\linewidth]{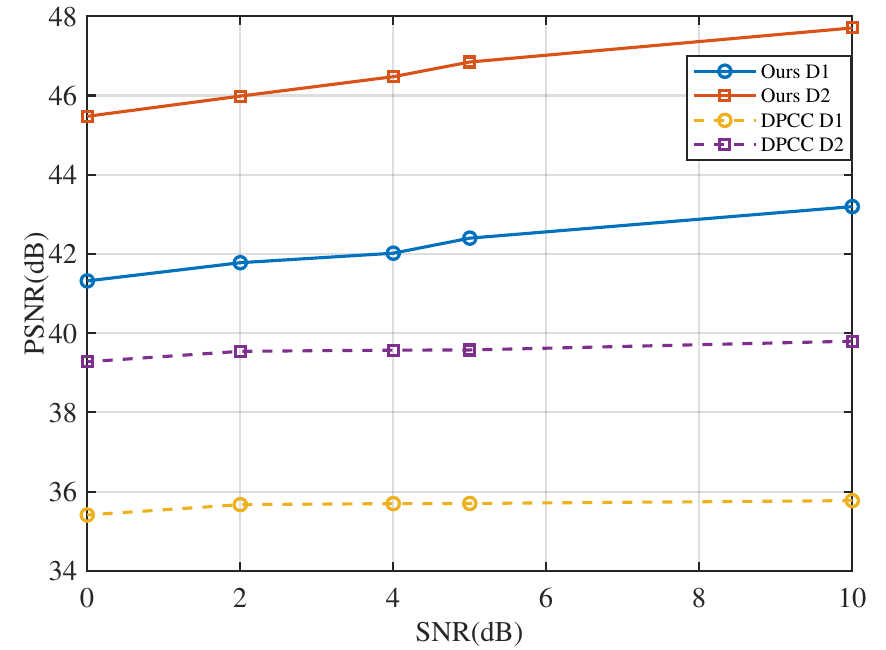}
\caption{ Comparison between the reconstruction performance of our model and DPCC under different SNR.}
\label{fig:SNR}
\end{figure}

\begin{figure}[]
\centering
\subfloat[Ours]{\includegraphics[width=2cm]{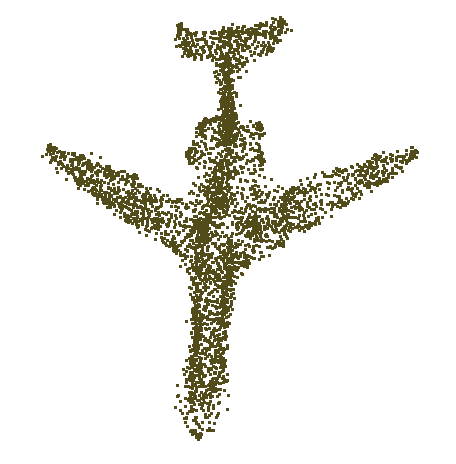}} 
\subfloat[DPCC]{\includegraphics[width=2cm]{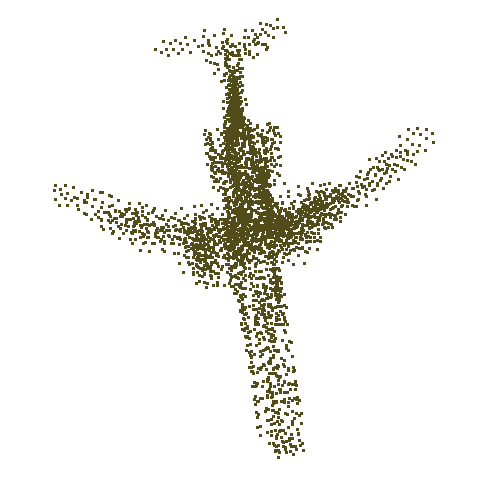}}
\subfloat[G-PCC]{\includegraphics[width=2cm]{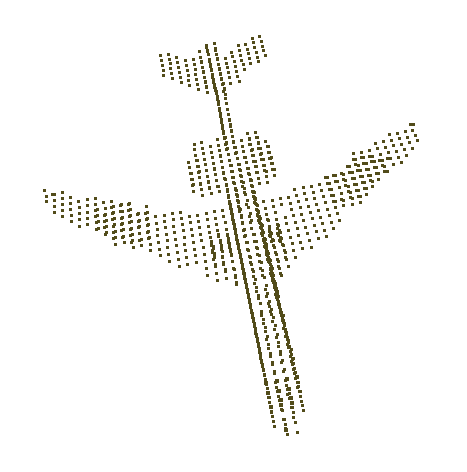}}
\caption{Visualization results of the three models. } 
\label{fig:visualization}
\end{figure}

\subsection{Experimental Setup}

We use point cloud data from ModelNet40, consisting of $9,843$ training samples and $2,468$ testing samples. Each point cloud is downsampled to $N=8,192$ points using the FPS algorithm. For compatibility with benchmark methods, we scale the data to the range $[0, 63]$ for input and normalize the output for performance evaluation. We consider an AWGN channel and test under the noise conditions described in Section \uppercase\expandafter{\romannumeral2}. For local information extraction, we use the Adam optimizer with an initial learning rate of $0.0005$ and a batch size of $24$. For global information extraction, we follow the MVTN settings \cite{MVTN}. In the second stage, since the training is more focused on local information extraction, we maintain consistency with the first-stage settings. To evaluate reconstruction performance across different SNR levels, we set the number of patches $S=64$, local semantic information $d=8$, projection maps $w=4$, and extracted global information $D'=4$. For compression performance evaluation, $w$ and $D'$ remain fixed while $S$ and $d$ vary. In ablation experiments, we use $S=64$, $d=8$, and $w=4$. We use D1 and D2 in Section \uppercase\expandafter{\romannumeral2} as performance metrics.

\subsection{The Reconstruction Performance}

In this subsection, we use the state-of-the-art deep learning-based DPCC \cite{DPCC} and the standard octree-based MPEG G-PCC \cite{GPCCTEST} as benchmark compression methods. We also employ the lossless transmission model from Section \uppercase\expandafter{\romannumeral3} to calculate transmission volume, assuming 16-bit precision for each parameter. It should be noted that our model need account for the transmission volume of local semantic information.

For G-PCC, we set initial parameters based on general test conditions \cite{GPCCTEST} and vary quantization parameters to obtain different rate-distortion points. DPCC's compression performance is adjusted by modifying its bottleneck size.

We compare the reconstruction performance under different compression settings, with results shown in Fig.~\ref{BPP}a and Fig.~\ref{BPP}b. Although our model is more sensitive to compression than DPCC, it achieves a 4 to 6 dB higher reconstruction quality. The compression performance of our model is influenced by the patch centroid coordinates $(S,3)$ we transmit, which are crucial to our patch-based approach. Any offset in these coordinates can significantly degrade performance, especially at low BPP. We also test DPCC whose centroid coordinates are generated by prediction. For DPCC, the reconstruction quality improvement is within 1dB even we send receiver precise centroid coordinates. The results show that our model outperforms DPCC, despite the need to transmit patch centroid coordinates.

Particularly, we also investigate how SEPT \cite{SEPT} performs under two channel conditions. Based on different applications, SEPT is a point cloud transmission scheme with good compression capability, while our scheme focuses more on high reconstruction quality. SEPT cannot simply increase transmission volume to significantly enhance reconstruction in certain scenarios.

To ensure fairness, we keep our model parameters fixed across different channel conditions. By adjusting the bottleneck size in DPCC under varying SNRs, we maintain comparable compression levels between DPCC and our model.

\begin{table}[tbp]
\centering
\caption{Model size and encoding\&decoding time results}
\begin{tabularx}{\linewidth}{|X|X|X|}
\hline
Method & Model size(MB) & Use time(s) \\ \hline
Ours   & 63.46          & 0.085219             \\ \hline
DPCC   & 416.38         & 0.150613             \\ \hline
G-PCC  &    N/A            & 0.052727             \\ \hline
\end{tabularx}
\label{size and rate}
\end{table}

The experimental results under different channel conditions are shown in Fig.~\ref{fig:SNR}. Our model degrades by approximately 2 dB in severe channel conditions, demonstrating better robustness against channel noise compared to DPCC, due to superior feature extraction. Notably, the DPCC model's bottleneck size remains relatively constant, which explains its stable performance.

We also present the reconstructed point cloud visualizations of the three methods at SNR $=5$ and BPP $\approx 0.6$ in Fig.~\ref{fig:visualization}, demonstrating that our model outperforms the others. G-PCC, in its standard configuration, produces fewer points on complex models due to the tree depth limitation.

\subsection{Model Size and Encoding\&Decoding Speed }

In this section, we discuss the model size, encoding, and decoding speed of our point cloud semantic transmission system, comparing it with DPCC and G-PCC. Under the conditions of SNR $= 5$ and BPP $= 0.6$, we measure the average transmission time for test files $63^{rd}$ to $127^{th}$ to accurately reflect the system's speed.

In Table \ref{size and rate}, we compare the speed and model size of various models. Our model processes each patch individually, resulting in a much smaller network scale and lighter weight compared to DPCC. Thus, our method achieves 56\% faster processing times than DPCC, thanks to the network's simplicity. Meanwhile, although G-PCC is slightly faster than our model, it sacrifices reconstruction quality.

In summary, our system achieves high-quality point cloud reconstruction at a fast operation speed. Additionally, its lightweight architecture allows it to run efficiently on devices with limited resources and computing power.

\subsection{Ablation Studies}

In this section, we present an ablation study to demonstrate the superior performance of our joint network compared to a single network. Additionally, we compare the two-stage training model with the no pre-training model to evaluate our method's performance.

\begin{figure}[tbp]
\centering
\includegraphics[width=0.9\linewidth]{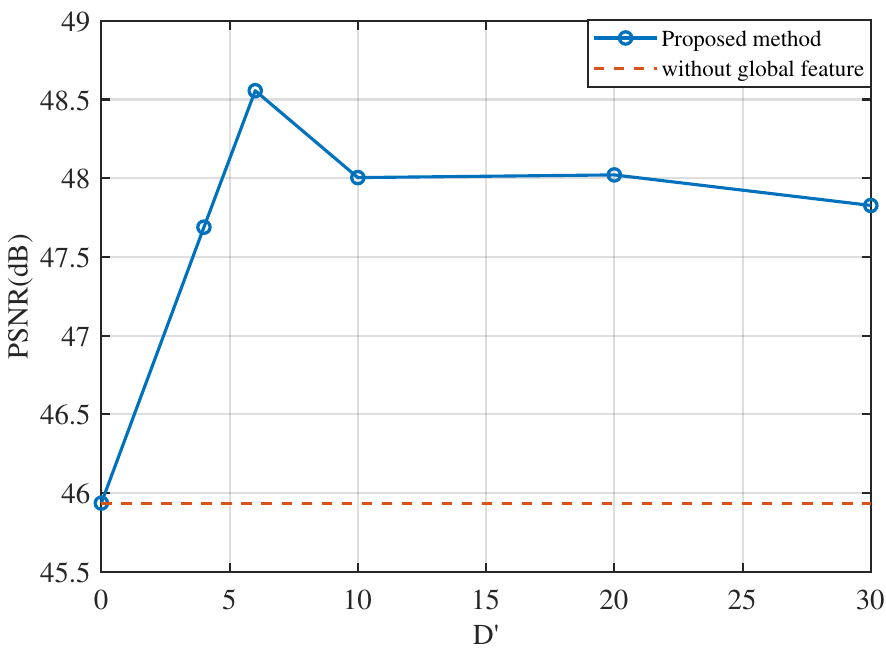}
\caption{Reconstruction performance under different global feature information. Here we use D2 as the indicator.}
\label{fig:photo_ablation}
\end{figure}

\begin{table}[]
\centering
\caption{ D2 PSNR results for two training strategies}
\begin{tabularx}{\linewidth}{|X|X|X|X|}
\hline
SNR(dB)               & 10 & 5  & 0  \\ \hline
Ours   & 47.705 & 46.845 & 45.474 \\ \hline
No pre-train  & 47.677 & 46.078 & 43.788 \\ \hline
\end{tabularx}
\label{strategy}
\end{table}

We first varied the global feature dimension $D'$ obtained by the global semantic encoder, with $S=64$, $d=8$, and tested reconstruction performance at SNR $=10$. The results, shown in Fig.~\ref{fig:photo_ablation}, indicate that incorporating global feature information significantly enhances reconstruction performance compared to excluding it. Optimal performance is achieved around $D'=6$. Notably, even small-scale global information can greatly improve reconstruction. However, the optimal $D'$ must balance with local feature size $d$. Excessive global information can degrade detail reconstruction, leading to performance fluctuations

We compare the reconstruction differences between the two-stage training model and the no pre-training model under different SNR conditions, as shown in TABLE \ref{strategy}. The performance of the no pre-training model drops significantly at low SNR. These results highlight the necessity of our two-stage training approach for the system. When noise is directly involved in training, especially under severe channel conditions, the lightweight semantic information extraction system becomes highly susceptible to noise, leading to a reduced information extraction capability.

\addtolength{\topmargin}{0.03in}

\section{Conclusion}

In this paper, we proposed a point cloud transmission system designed for high reconstruction quality. The system integrates point-based and graph-based neural networks to enhance the extraction of point cloud semantic information through multi-dimensional processing. A two-stage training strategy and layered transmission are employed to optimize performance for the physical channel environment. The point cloud is then reconstructed using a semantic decoder at the receiver. Simulation results validated the performance of the proposed SemCom method for point cloud transmission.

\vspace{12pt}

\end{document}